\documentclass[a4paper,12pt]{article}

\usepackage[dvipsnames]{xcolor}
\usepackage{amsmath}
\usepackage{amssymb}
\usepackage{dsfont}
\usepackage[english]{babel}
\usepackage[utf8]{inputenc}
\usepackage{times}
\usepackage{graphics}
\usepackage{graphicx}
\usepackage[T1]{fontenc}
\usepackage[official,right]{eurosym}
\usepackage{url}
\usepackage{eso-pic}

\title{Is there an ethical Operational Research?}
\author{Odille Bellenguez$\dag$, Nadia Brauner$\ddag$, Alexis Tsoukiàs$^*$ \\ $\dag$IMT Atlantique, LS2N \\ $\ddag$Université Grenoble Alpes, CNRS, Grenoble INP, G-SCOP \\ $^*$CNRS-LAMSADE, PSL, Université Paris Dauphine}
\date{}

\begin{document}

\thispagestyle{empty}

\enlargethispage*{8cm}
 \vspace*{-38mm}

\AddToShipoutPictureBG*{\includegraphics[width=\paperwidth,height=\paperheight]{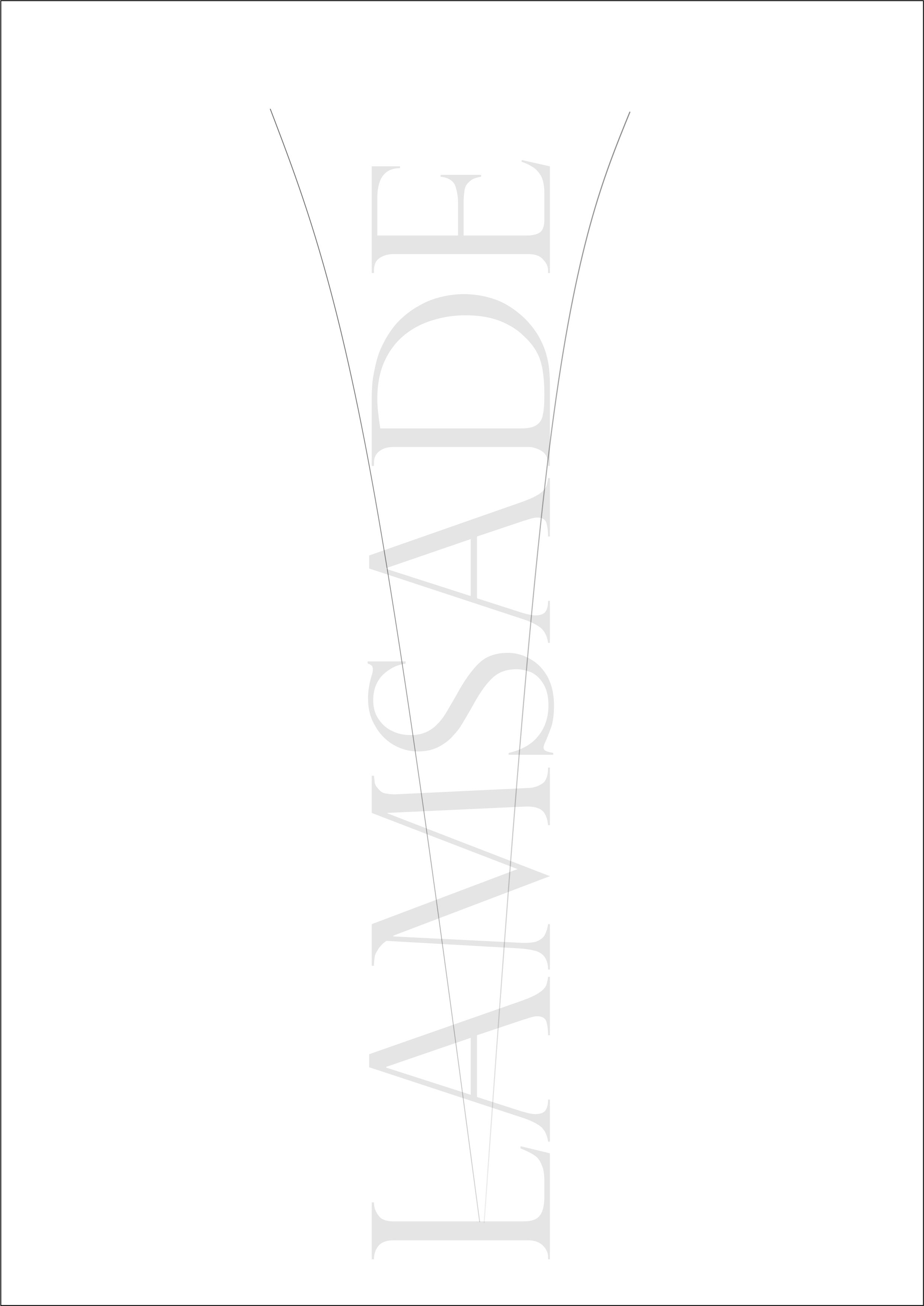}}

\begin{minipage}{24cm}
 \hspace*{-28mm}
\begin{picture}(500,700)\thicklines
 \put(60,670){\makebox(0,0){\scalebox{0.7}{\includegraphics{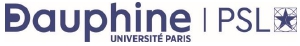}}}}
 \put(60,70){\makebox(0,0){\scalebox{0.3}{\includegraphics{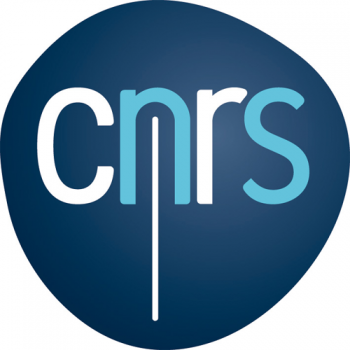}}}}
 \put(320,350){\makebox(0,0){\Huge{CAHIER DU \textcolor{BurntOrange}{LAMSADE}}}}
 \put(140,10){\textcolor{BurntOrange}{\line(0,1){680}}}
 \put(190,330){\line(1,0){263}}
 \put(320,310){\makebox(0,0){\Huge{\emph{403}}}}
 \put(320,290){\makebox(0,0){June, 2022}}
 \put(320,210){\makebox(0,0){\Large{Is there an Ethical Operational Research?}}}
 \put(320,100){\makebox(0,0){\Large{Odile Bellenguez, Nadia Brauner, Alexis Tsoukiàs}}}
 \put(320,670){\makebox(0,0){\Large{\emph{Laboratoire d'Analyse et Mod\'elisation}}}}
 \put(320,650){\makebox(0,0){\Large{\emph{de Syst\`emes pour l'Aide \`a la D\'ecision}}}}
 \put(320,630){\makebox(0,0){\Large{\emph{UMR 7243}}}}
\end{picture}
\end{minipage}

\newpage

\addtocounter{page}{-1}

\maketitle

\abstract{The ethical dimension of Operational Research and Decision Aiding, although not a new subject, turns to be a concern, both for the large public and the OR community, because of the wide spread of autonomous artefacts endowed with decision capacity thanks to the use of models, methods and tools developed within our field. The paper addresses the question of whether there exists an ``Ethical Operational Research'', identifies the ethical questions which are specific to our professional community and suggests research topics which, although independently developed, are relevant for handling such questions.}

\newpage


\section{Introduction}

There is an increasing interest and discussion about ``Ethical Operational Research'' and more generally about ``Ethical'' or ``Responsible'' Decision Support (see \cite{Tsoukias2021}). The topic is not really a new one: there is a EURO Working Group on Ethics\footnote{See \url{https://www.euro-online.org/websites/ethicsandor/}} since 2002 and there is a ``Prometheus Oath'' written by J.P. Brans, founder of this Working Group (see \cite{Brans2002b}). Interested readers can see two excellent surveys in \cite{Brans:2007} and \cite{OrmerodUlrich2013} of the literature on this topic.

Under such a perspective this contribution just continues an ongoing discussion already started in the 60s and early 70s (see \cite{Ackoff1974}, \cite{Churchman1968}) and continued since then (see \cite{gallo2004}, \cite{gass2009}, \cite{Gass1994}, \cite{LeMenestrelVanWassenhove2009}, \cite{Wenstop2010}). The reason for which these topics turn to be discussed is related to the increasing diffusion of ``autonomous artefacts'' with augmented decision capacity. Both the wide public, but also scientists and policy makers are concerned by the wide spread of devices and processes which decide or recommend decisions using ``algorithms'' or ``methods'' which are felt to be non-controllable, dangerous, biased, unfair, inexplicable with unknown long-term impacts (\cite{STUD2019}).

We share such concerns with a near discipline, Artificial Intelligence\footnote{See the topics discussed at the conference:
\url{https://facctconference.org/} or the Mechanism Design for Social Good Working Group: \url{https://www.md4sg.com/}. See also the High-Level Expert Group for Trustworthy Artificial Intelligence recommendation to the European Commission: \url{https://digital-strategy.ec.europa.eu/en/library/ethics-guidelines-trustworthy-ai}.} and more precisely with Computational Social Choice. Most of these concerns are not really new (see the discussion in \cite{Tsoukias2021}), but it pays to continue the discussion as we do with this paper.

The paper is organised as follows: in section \ref{setting} we introduce a general framework for the paper discussion. In section \ref{Forwhom} we identify the type of ``ethical questions'' that could be of interest for this paper. In section \ref{profethics} we discuss the questions which can be raised within our professional practice as decision analysts (or designers of decision support devices). In section \ref{researchethics} we briefly detail some research topics which although stand alone are at the same time useful in order to improve how we handle the topics discussed in section \ref{profethics}. We summarise the discussion in section \ref{conclusions} where we ultimately provide a reply to the question in the title.

\section{Setting} \label{setting}

In order to frame our discussion we are going to use a precise setting introduced and discussed in \cite{MeinardTsoukiasEJOR2018}, \cite{MeinardTsoukias2022} and \cite{Tsoukias07aor}.

We consider a situation where a ``client'' (an entity implied in a decision process), asks an advice to a ``decision analyst'' about how to improve his/her conduct with respect to that decision process. We also consider situations in which the ``client'' does not seek an advice for a precise decision process, but for a class of decision processes, the advice coming under the form of a device/system/software which is supposed to elaborate recommendations; the decision analyst in such cases being the designer of such system.s. Under such a perspective the decision support activities can be seen as: \\
 - either the direct interactions between client and analyst in order to elaborate a recommendation; \\
 - or the design of an appropriate system which on its turn will compute or help to compute a recommendation. \\
Clearly many combinations between these two extreme cases are possible in the real world.

As already partially mentioned in \cite{Tsoukias08ejor}, the use of a formal decision aiding methodology implies considering three different dimensions.

\begin{enumerate}
  \item An axiomatic dimension, establishing the conditions under which it is possible to use protocols/algorithms/models in a meaningful and useful way.
  \item An algorithmic dimension, considering the size of the solution space, the necessary data (availability, accessibility, storage, quality) as well as the necessary computing resources.
  \item A pragmatic dimension related to the conditions under which a decision aiding process is valid and legitimate (see \cite{LandryBanvilleOral96}, \cite{LandryMalouinOral83}, \cite{MeinardTsoukias2022}).
\end{enumerate}

\section{Ethics for whom?} \label{Forwhom}

First of all we need to identify different categories of concerned individuals. To be more precise: as Operational Researchers or Decision Analysts we may raise ethical questions for different purposes and under different perspectives. Not all of them are necessarily of interest for a scientific investigation.

\vspace{5mm}

We maybe raise ethical questions just because we are conscious citizens. These are the typical questions which all of us some day need to consider, but generally are related to our own individual values. These cannot be matter of study, analysis or guidelines and principles of conduct within our discipline if not respecting the very general values our societies consider relevant. But there is nothing specific to the fact that we are decision analysts.

Just to be more precise: accepting to provide support and models for military operations can raise ethical and/or moral questions to any among us, but the positive or negative reply is a matter of personal choices. Some of us will be happy to do it, others not and others might be indifferent. We cannot establish any ethical guideline on how to handle such issues.

\vspace{5mm}

Some among us, besides being citizens are also teachers or researchers (in OR). Independently from our specific discipline and research field, there exist ethical questions related to our precise role of scientists. Such ethical questions concern our conduct as teachers (with respect to our students and colleagues) and researchers (with respect to our near scientific communities and the science in large).

Such issues are generally handled through deontology charts (how to behave appropriately with the students, how to conduct experiments, how to write papers, how to quote the existing literature etc.), or specific debates in philosophy of sciences (see e.g. \cite{Coutellec}), but are not specific to our discipline and research areas (see e.g. the Practical guide “Integrity and responsibility in research practises” from the CNRS ethics committee \cite{cnrs}, the Singapore Statement on Research Integrity \cite{singapore}, the OECD Best Practices for Ensuring Scientific Integrity and Preventing Misconduct \cite{OECD}, The European Code of Conduct for Research Integrity by the European Academies \cite{ALLEA}).

There are instead two areas of ethical concerns which are specific to our domain and role of Decision Analysts. These are related to our profession (providing advice to decision makers or designing tools to be used by decision makers) and to our research in the broad area of Operational Research and Decision Analysis.

\section{Professional Ethics} \label{profethics}

It is interesting to note that large part of the debate and the literature about the ethical dimension of our discipline and the practice related to it, originate from discussions about professionalism and deontology in our profession started at the late 60s. At that time the idea of creating a professional body of ``chartered'' operational researchers or decision analysts (later on called ``OR Fellows'' by the ORS in UK) ignited a debate which lasted several decades (and still persists) before reaching any practical conclusion (\cite{Ackoff79a}, \cite{Ackoff79b}, \cite{Churchman1970}, \cite{Kirby06}, \cite{ORSA71}, \cite{RosenheadMitchell1986}). This discussion moved beyond the UK and USA professional bodies (see \cite{Rauschmayeretal2009}) and we can try to summarise under a specific perspective (the one of conducting rationally decision aiding processes).

As decision analysts we provide support to clients. We are not the only professionals who provide decision support: lawyers, accountants, physicians, psychotherapists, engineers, just to give some examples, help their clients to handle their problems and they do so using some scientific knowledge and approach, thus distinguishing their profession from just informal intuitive advice to friends and relatives. There are two topics we need to consider here: \\
 - What does make our decision support different from other equally scientifically based decision support activities? In other terms why decision analysts are not psychotherapists? \\
 - Since we nevertheless share some features with other professions, who already considered the problem of deontology, compliance, unsatisfied clients, young professional training etc., why our profession should not establish similar protocols, practices, training modules etc.?

\vspace{5mm}

We are not going to pursue further this discussion because it is out of the scope of this paper, but we can identify the ethical problems to handle within our profession: under which conditions we can claim that our professional advice to a client satisfies appropriate ethical standards and who establishes such standards?

We need to separate two distinct cases raising potentially different types of ``ethical questions''. The first one is the case where analysts directly provide at a client some advice on how to handle a problem within a decision process, a typical case being organising the shifts of the personnel at the emergency department of a hospital or managing a large call for tenders for software COTS for an IT industry.

The second case consists in designing generic methods, protocols or software aimed at being used for a precise class of decision problems, possibly customising such products for and with specific clients. Typical examples here include, supply management packages, flow-shop scheduling procedures, generic recommender systems etc.

There are certainly mixed cases as well as cases where specific applications become generic ones (a well known example being the yield management procedures originally designed for a specific client and then developed as stand alone customisable methods; see \cite{Smithetal1992}).

Despite the apparent differences between these two cases we will develop a unique argumentation based on two aspects: \\
 - the use of a decision aiding methodology; \\
 - the unveiling of hidden or implicit hypotheses and assumptions within our models and/or methods.

\subsection{Decision Aiding Methodology}

We start considering our profession as being characterised by the use of a formal language, the pretention of using/introducing a rationality model within the client's decision process and the use of algorithms (see \cite{Tsoukias08ejor}). As introduced in \cite{Tsoukias07aor}, a decision aiding process can be defined as the interactions between a ``client'' (asking for advice) and an ``analyst'' (providing the advice) and can be represented through a set of ``deliverables'' such as: \\
 - a description of the problem situation; \\
 - a problem formulation; \\
 - an evaluation model; \\
 - a final recommendation.

In constructing such deliverables we make choices (as analysts). For instance: \\
 - in order to represent the likelihood of an event we may adopt a probability measure (while other measures are possible); \\
 - in order to compute a majority for a voting procedure we may adopt the ``Borda'' rule (but others are possible); \\
 - in order to model the impact of the combined realisation of some decision variables we assume this impact being linearly defined with respect to the variables (but other choices are possible). \\
Such ``technical'' choices are most of the time uncontrollable by the client and we (the analysts) are the only able to measure the consequences and to guarantee the meaningfulness and usefulness of their use.

We also do further hypotheses which are less technical, but nonetheless important. For instance: \\
 - in calculating the economic impact of a given infrastructure we consider a ``territory'', but how this territory is chosen/defined? \\
 - in designing a supply chain we consider the client's costs and time constraints, but are these the only constraints we should take into account? \\
 - in order to set up a vendor rating procedure should we compare the suppliers between them or only with respect to quality standards? \\
Despite such choices being agreed with the client(s), it is unlikely the client(s) really realise the extension to which modifying any of these hypotheses can modify the outcome of the decision aiding process. In other terms, beyond any generic deontological constraint due to the fact that we have a professional relationship, we need to consider the specific constraints our condition sets. We can try to summarise these through the following points.

\begin{itemize}
  \item \emph{Are we sufficiently critical?} The fact we work for a client does not mean we cannot or we should not have a personal and independent perspective about what the client claims to be the problem to consider. We need to be able to show to the client aspects of the problem or other problems she does not see. At the same time we should be ready to modify our perspective and learn from the client's claims, values and beliefs.
  \item \emph{Where does rationality come from?} This topic is extensively discussed in \cite{MeinardTsoukiasEJOR2018}. The point to raise in this discussion is that there are several sources of ``rationality'', from external norms and standards to subjective behaviours and argued beliefs. What we need is understanding which among such different sources we use with that precise client for that precise decision process we are involved to. The pretention of introducing one or more dimensions of rationality to the client's reasoning for her problem, is what, most of the times, legitimates our action as analysts. Having a clear idea about where such dimensions of rationality come from is essential in establishing an appropriate professional relationship with the client.
  \item \emph{Can we explain, justify and easily revise or update?} Most of the times we deliver both a model and the results of applying a set of methods to the constructed model. This implies choosing among what our technical knowledge and skills provide, following what the client claims being her problem. However, we do many technical choices which not always are ``obvious'', at least for the client and/or the other involved stakeholders. It might not be always necessary, but we need to ask ourselves: should I be asked, am I able to explain why we did such a technical choice, to completely justify such a choice against an appeal to a court and to defend the choice against an ``adversary'' analyst? Moreover, since modelling for decision support purposes is always a learning process, we also need to ask ourselves: how easy is it to revise and update the model and the methods in case the data change, the values and the opinions change, the problem setting (and formulation) can change.
  \item \emph{Is the result convincing for us, for the clients and for the involved stakeholders?} Providing decision support means constructing convincing arguments for some potential action to undertake. Such convictions concern three different categories of stakeholders. We first need to convince ourselves that our advice is sound with respect to our technical knowledge and our methodology. We then need to convince the client that our advice is appropriate with respect to the problem the client has, the decision process for which the advice has been asked: the client needs to feel owner of the advice received. Finally we need to convince the rest of the stakeholders that the advice to the client was legitimately designed, that we have been critical and that the impacts of our advice being adopted have been understood.
\end{itemize}

\subsection{Hidden hypotheses}

Although in a professional setting we deploy a formal decision aiding methodology, we are always induced in assuming a number of hypotheses as granted or given. Some of these can become explicit through an appropriate use of our methodological knowledge. The fact that using a linear (additive, separable) utility function in order to aggregate the impacts of decisions along different attributes implies assuming that such impacts are commensurable and can be traded among them, is part of our methodological knowledge which will consistently impede us to use such a method in case this hypothesis does not hold (i.e. the client does not accept it). However, there are potential misuses and errors which can occur under certain implicit hypotheses, for which we are not really trained and prepared to handle. Such implicit assumptions (often ignored exactly because implicit) need to be explicitly identified and handled if we want to build a trusted relationship with our clients or in case we want a trustworthy use of any autonomous artefact using our models and methods.

   \begin{itemize}
    \item \emph{Cognitive biases.} Decision Analysts are subject to the same cognitive limitations as other humans. They have personal values, personal cognitive limits, personal habits, culture and feelings. How much these influence the way through which the client's problem and the information provided are modelled? Other professional bodies impose specific training in order to handle such questions or establish specific protocols and external assessors to be used by those clients who may doubt about the analysts' biases. It is certainly true that our profession does not consider any specific training or appealing procedure, although these problems exist and should at least be discussed with our clients.
    \item \emph{Exceptional circumstances.} Consider a classic risk management model and a situation characterised by extreme risks (extremely unlikely to occur events, but with extremely high impacts in case they actually occur; see \cite{Nott2006}). Would our model still work appropriately? Consider a model of extreme risk theory: would this still apply for emerging risks management (see \cite{Bier2016})? Consider a supply chain model. Would this work and be robust under any possible circumstances? And would that model still apply if the supply chain problem concerns humanitarian logistics (see \cite{TomasiniWassenhove2009})? \\
    The above examples are just cases where mainstream methods and models have been proven to be inappropriate when exceptional circumstances occur. This leads to a general question: given any method or model we suggest using for advising a client, will this advice still hold if such exceptional circumstances occur and if not what do we suggest the client to do? Which raises the question of whether we know the application limits of generalisation of whatever we suggest as recommendation to our clients.
    \item \emph{Data}. All our methods and algorithms require data. Not only the ones provided by the client, but also data about the ``territory'', the ``landscape'', the ``culture'',  the ``economical and social context'', the ``organisation'' where the decision process for which our help is requested is going to be used. Data are collected, stored, transferred, transformed, manipulated, along ``pipelines'' which are far from being with no impact upon the final outcome. Moreover, data, although they belong to nobody, can be protected by ``rights'', private or collective, social, economic or cultural. In designing methods and models we need to consider both the data pipeline quality as well as the rights protection issue (see \cite{Christophidesetal2021}).
    \item \emph{Algorithms}. Most of the times our methods require efficient algorithms. Most of the times we need to trade-off between efficiency and accuracy or even optimality. Most of the times we also need to take into account other features of the algorithms such as manipulability, strategic proofness, security, robustness to adversarial attacks, black-box effect etc. Most of the times our clients are not aware of what is the impact of choosing an algorithm instead of another. Clients are also usually unaware of the software differences when algorithms are coded and of the computing resources necessary to run them. It is unlikely our clients will ever be tempted to learn all such topics, but is our ``ethical'' obligation to know them and let the client understand which are the stakes at play when choosing a precise algorithm and a precise software implementation.
    \item \emph{Impossibilities}. Not all methods fit to any type of decision problem. Generally speaking we know that most of the times, given a set of properties to be satisfied by the outcome of a potential algorithm, these are inconsistent. In other terms there is no algorithm able to satisfy at the same time all the desired properties (see \cite{arrow1book51}, \cite{Brandtetal2016book}, \cite{vinck1theo92}). This is not really a problem in terms of computing solutions, but we need to know which properties are satisfied by which algorithms and we need to be able to explain that to our clients. In other terms we need to be able to explain to our clients what an impossibility theorem means for her problem and which are the different partial solutions we can adopt (and at which ``price'' in terms of satisfied properties).
    \item \emph{Long term consequences}. When American Airlines started studying yield management methods in order to manage the ticket pricing (see \cite{Smithetal1992}), nobody (within the company and in the broad Operational Research community) could ever imagine the impact these methods will have on the travelling industry and the travelling habits within our societies. Today potentially any operator running a travelling business (including trains and buses) uses a yield management method in order to price dynamically tickets. The whole industry in this field changed its business model and each single consumer modified its willingness to pay for a travelling ticket (independently from business or leisure travelling). We are not going to discuss here whether this had a long term positive or negative impact, although some may discuss the consequences on the tourism industry, the house renting industry, the environmental impact etc. The ``ethical'' question is that all such impacts have never been discussed neither within the company nor within the society. Nobody anticipated, discussed or even questioned the new business model underestimating the impact of a simple optimal pricing method.

    Providing models, methods, tools, aiding to improve decision making has impacts which can go far beyond the client and the other involved stakeholders. Most of the ethical questions we rose in the previous paragraphs are related to a precise decision aiding process, the actors involved and the immediate impacts. But we need to also consider impacts which will occur on a long, very long term and for stakeholders, citizens, territories and biomes who never ever thought that changing the optimal production policy of a company could change their lives half a century later.
   \end{itemize}

\section{Ethics in OR research} \label{researchethics}

We have seen that most of the ``ethical questions'' about our discipline concern its use in the real life and the way through which we handle the relations with our clients, the relevant stakeholders and the use of our decision aiding methodology.
The reader may note that many of such questions are related to topics addressed in our research (mostly indirectly) already since the 70s such as the axiomatic analysis of voting procedures, the analysis of behavioural biases or the development of problem structuring methods. The question we raise at this point is: \emph{``are there specific research topics in Operational Research related to our ethical questions''}? Being more precise: probably any research topic in our discipline could be relevant for our ethical questions, but are there some new or more relevant ones?

\begin{enumerate}
  \item \underline{\textbf{Are we aware?}} The first class of research questions concerns awareness. In section \ref{profethics} we raised several ethical questions concerning the use of models, methods, algorithms, protocols, etc. The fundamental remark is that none among such tools (which we use in order to advise our clients in their decision processes) are ``neutral''. Using one instead of another can have short or long term consequences which are independent from what the client asks or the situation requires. The question is: are we aware of such consequences? And the consequent research question is: Do we know how to choose appropriately our tools? In other terms, do we know what each tool can do, cannot do, the conditions under which they can be used and provide meaningful results?

Axiomatic characterisations, representation theorems, simulations, are certainly research fields in our discipline which provide results usable in order to reply to the above demands. Under such a perspective we may emphasise that using numerical simulations and experimental settings (testing protocols, behavioural biases and modelling hypotheses) result in very useful tools helping to undercover hidden behaviours and tacit assumptions which could be concealed during the use of any among our tools.

We may also emphasise that the more we use the ``open science'' paradigm, sharing data, software codes, results and findings, greatly improves our capacity to increase awareness about what, when and how works (or does not work).

  \item \underline{\textbf{Do we help others becoming aware?}} Being ourselves (as analysts) aware of what our tools and our methodology can (and cannot) do is necessary, but not sufficient. Both our clients and the involved stakeholders (possibly the society as a whole?) need to develop awareness of what our tools can (and cannot) do.

  This of course raises a far larger topic about how scientists can and should communicate their findings to a ``non-scientific'' audience in a way that increases and strengths awareness and autonomy (see \cite{Baueretal2007} or \cite{BucchiTrench2014}), but remains relevant for our discipline. Large part of the general public gets used in misusing statistics and other quantitative information, in using inappropriately averages, indexes, protocols and codes, producing totally meaningless results and conclusions (see \cite{Gigerenzer02} or for more fun \cite{Barker2020}). Even large and prestigious institutions, not to talk about public agencies and governments misuse such information in order to justify regulations and policies (for a famous example see the incredible diffusion of a meaningless index such as the Shanghai ranking of the Universities: \cite{Billautetal2010}). Under such a perspective it pays learning to use simple heuristics facilitating communicating quantitative information in a meaningful way (see \cite{GigerenzerTodd99}).

At the same time developing general frameworks which allow to unify a field of models and provide a unique frame within which interpret, explain and justify methods and protocols helps increasing awareness for any stakeholder (and the general public) involved in a decision process. The reader can see the impact of measurement theory, \cite{roberts79}, in establishing a rigorous notion of meaningfulness: \cite{NarensLuce1990} \cite{Roberts1980}, or the impact of conjoint measurement theory in unifying the field of multiple criteria decision analysis, \cite{BouyssouPirlot2009book}.

Last, but not least, assuming a problem driven decision support attitude, instead of a method driven one, generally allows to improve communication with the client and enhance awareness about why certain methods will not fit in that precise problem situation, while others might be more appropriate. The result is adopting an ``horizontal'' or ``methodological'' view of our discipline and not seeing Operational Research as just a collection of methods (see \cite{Tsoukias08ejor}).

  \item \underline{\textbf{Are we critical?}} A decision aiding process is certainly a set of activities involving the client (who asks for advice) and the analyst (who provides the advice). However, we already observed that the choices we do, while conducting the decision aiding process, have impacts far beyond these two stakeholders. Moreover, remaining confined within the client's demand and/or the analyst's perception can result in missing other opportunities the decision aiding process offers.

  Most problem structuring methods (see \cite{Keeney92}, \cite{rosen1book89}) emphasise that decision problems are constructed (and not identified) and that solutions critically depend on how problems are formulated. More recently the problem of constructing the set of alternatives (which is at the centre of the decision model) has turned to be a research topic (see \cite{ColorniTsoukias2020}, \cite{FerrettietalEJOR2018}).

  Under such a perspective, research in the following three areas is extremely important in order to improve and expand our capacity to interact with the clients and the other stakeholders developing our critical view of the decision aiding process: \\
       - \textbf{problem structuring methods} in general, since they provide a general framework for supporting the whole decision aiding process (\cite{Rosenhead2006}); \\
       - \textbf{design theory} as a formal tool for developing ``out-of-the-box'' alternatives beyond the dominant designs usually suggested as solutions (\cite{Alexander1982}, \cite{LeMasson2013}); \\
       - \textbf{preference learning} because whatever we use in order to elaborate an advice is learned through interaction with the clients and/or accessing data and information (\cite{FurnkranzHullermeier2010}).

  \item \underline{\textbf{Do we help others becoming critical?}} Keeping a critical perspective with respect to the problem situation as it appears to be (or as the client makes it appearing) is certainly important for our ethical questions. However, it is not sufficient. The client and the involved stakeholders also need to develop a critical perspective about what happens both within the decision process they are involved in and the decision aiding process.

Such a process is very much a matter of ``convincing'': first of all ourselves (we remain within standards of meaningfulness), then our clients (they get something they feel it helps them) and then the rest of the world (the whole process was legitimate; see \cite{MeinardTsoukiasEJOR2018}, \cite{MeinardTsoukias2022}).

Under such a perspective our clients and the other stakeholders need to be able to reply positively to the question: is this advice going to resist to any arguing against it, arguing grounded on data, procedures, protocols, and authority? A formal framework where such problems are discussed is formal argumentation theory (see \cite{AmgoudPrade2009}, \cite{Dung95}, \cite{Ouerdaneetal2010}, \cite{TrevorBench-Capon2003}).
\end{enumerate}

\section{Conclusions} \label{conclusions}

Let us summarise the discussion and the claims we introduced. Ethical questions arise in our everyday life as well as in our professional life independently from what our profession is. The questions we are interested in this paper derive from our specific profession as decision analysts. As a side effect we need to consider which research topics, while independently developed, can help us in handling such ethical questions.

Under such a perspective we need to remember that decisions (what we are supposed to help making) are value driven and not data driven: data are necessary, but not sufficient for making decisions. It is part of our profession to make understand these values, for us, our clients and the involved stakeholders. At the same time we need to remember that aiding to decide is problem driven and not method driven, which means we first need to understand the problem and then we need to think about solving it.

As we show in the paper, ethical questions have been introduced in our discipline since the very beginning. Our discussion emphasises two parallel issues we need to consider when we try to handle such questions. The first concerns awareness of what methods are, can do, cannot do and how choosing any among them is not neutral with respect to the solution computed and the recommendation provided. The second is the development of a critical attitude about the consequences of our modelling choices which goes beyond the usual relation analyst/client.

A first point to make is that there are no universal procedures, protocols, algorithms or methods. Any of them will unfit for some reason and for some kind of problem situation and for some type of client and we need to know how to handle this. Keeping a critical attitude with respect to the clients' demand and to our profession helps on a long run both our clients and our profession. There are many hidden hypotheses and implicit choices in modelling and solving a decision problem and these need to be clear to us (as analysts), to our clients and to the stakeholders who might be involved in the problem situation.

Is there an ethical Operational Research? If ethics consists in applying to our profession standards of morality (in whatever way these have been established) then our reply is negative. But if ethics consists in assuming our responsibility for what we offer to our clients then yes. Although we may not be liable for what our clients decide using our advice and/or our tools, we are responsible for many (avoidable) consequences which can occur. We have a power and we need to use it with responsibility.

\section*{Acknowledgements}

The paper reflects discussions hold within the Special Interest Group on Ethics in Operational Research of the CNRS funded National Research Network on Operational Research (GDR-RO, \url{http://gdrro.lip6.fr/}). We specially thank the participants to the workshop hold at the LAMSADE, Université Paris Dauphine the 26/11/2011, \url{http://gdrro.lip6.fr/?q=node/254}. Christine Solnon read an early version of this document and her comments greatly improved it. 


\begin{thebibliography}{10}

\bibitem{ALLEA}
All~European Academies.
\newblock The european code of conduct for research integrity, 2017.
\newblock
  https://allea.org/wp-content/uploads/2017/05/ALLEA-European-Code-of-Conduct-for-Research-Integrity-2017.pdf.

\bibitem{Ackoff1974}
{R.L.} Ackoff.
\newblock The social responsibility of operational research.
\newblock {\em Operational Research Quarterly}, 25:361--371, 1974.

\bibitem{Ackoff79a}
{R.L} Ackoff.
\newblock The future of operational research is past.
\newblock {\em Journal of Operational Research Society}, 30:93--104, 1979.

\bibitem{Ackoff79b}
R.L. Ackoff.
\newblock Resurrecting the future of operational research.
\newblock {\em Journal of the Operational Research Society}, 30:189--199, 1979.

\bibitem{Alexander1982}
{E.R.}. Alexander.
\newblock Design in the decision making process.
\newblock {\em Policy Sciences}, 14:279--292, 1982.

\bibitem{AmgoudPrade2009}
L.~Amgoud and H.~Prade.
\newblock Using arguments for making and explaining decisions.
\newblock {\em Artificial Intelligence}, 173:413–436, 2009.

\bibitem{arrow1book51}
K.J. Arrow.
\newblock {\em Social choice and individual values}.
\newblock J. Wiley, New York, 1951.
\newblock 2nd edition, 1963.

\bibitem{Barker2020}
H.~Barker.
\newblock {\em Lying Numbers: How Maths and Statistics Are Twisted and Abused}.
\newblock Little, Brown Book Group, London, 2020.

\bibitem{Baueretal2007}
M.~Bauer, N.~Allum, and S.~Miller.
\newblock What can we learn from 25 years of pus survey research? liberating
  and expanding the agenda.
\newblock {\em Public Understanding of Science}, 16:79 – 95, 2007.

\bibitem{Bier2016}
{V.M.} Bier.
\newblock {\em The Gower handbook of extreme risk : assessment, perception and
  management of extreme events}.
\newblock Routledge, London, 2016.

\bibitem{Billautetal2010}
{J.-Ch.} Billaut, D.~Bouyssou, and {Ph.} Vincke.
\newblock Should you believe in the shanghai ranking?
\newblock {\em Scientometrics}, 84:237 -- 263, 2010.

\bibitem{BouyssouPirlot2009book}
D.~Bouyssou and M.~Pirlot.
\newblock Conjoint measurement models for preference relations.
\newblock In D.~Bouyssou, D.~Dubois, M.~Pirlot, and H.~Prade, editors, {\em
  Decision Making Process}, pages 617 -- 672. J. Wiley, Chichester, 2009.

\bibitem{Brandtetal2016book}
F.~Brandt, V.~Conitzer, U.~Endriss, J.~Lang, and {A.D.} Procaccia.
\newblock {\em Handbook of Computational Social Choice}.
\newblock Cambridge University Press, Cambridge, 2016.

\bibitem{Brans2002b}
{J-P.} Brans.
\newblock {OR} ethics and decision: the {OATH} of {PROMETHEUS}.
\newblock {\em European Journal of Operational Research}, 140:191--196, 2002.

\bibitem{Brans:2007}
{J-P.} Brans and G.~Gallo.
\newblock Ethics in {OR/MS}: past, present and future.
\newblock {\em Annals of Operations Research}, 153(1):165 -- 178, 2007.

\bibitem{BucchiTrench2014}
M.~Bucchi and B.~Trench.
\newblock {\em Handbook of Public Communication of Science and Technology}.
\newblock Routledge, London, 2014.
\newblock 2nd edition.

\bibitem{STUD2019}
C.~Casteluccia and D.~{Le M\'etayer}.
\newblock Understanding algorithmic decision making: opportunities and
  challenges.
\newblock {EPRS} study, European Parlament, 2019.
\newblock 104 pages.

\bibitem{Christophidesetal2021}
V.~Christophides, V.~Efthymiou, {Th.} Palpanas, G.~Papadakis, and
  K.~Stefanidis.
\newblock An overview of end-to-end entity resolution for big data.
\newblock {\em {ACM} Computing Surveys}, 53:1 -- 42, 2021.

\bibitem{Churchman1968}
{C.W.} Churchman.
\newblock {\em Challenge to Reason}.
\newblock McGraw-Hill, New York, 1968.

\bibitem{Churchman1970}
{C.W.} Churchman.
\newblock Operations research as a profession.
\newblock {\em Management Science}, 17:B37--B53, 1970.

\bibitem{cnrs}
{CNRS ethical committee COMETS}.
\newblock Practical guide ``integrity and responsibility in research
  practises'', 2017.
\newblock https://comite-ethique.cnrs.fr/en/practical-guide/.

\bibitem{ColorniTsoukias2020}
A.~Colorni and A.~Tsoukiàs.
\newblock Designing alternatives for decision problems.
\newblock {\em Journal of Multi-Criteria Decision Analysis}, 27:150 -- 158,
  2020.

\bibitem{Coutellec}
L.~Coutellec.
\newblock Penser l'indissociabilité de l'éthique de la recherche, de
  l'intégrité scientifique et de la responsabilité sociale des sciences.
\newblock {\em Revue d'anthropologie des connaissances}, 13, 2:381 -- 398,
  2019.

\bibitem{Dung95}
{Ph.M.} Dung.
\newblock On the acceptability of arguments and its fundamental role in
  nonmonotonic reasoning, logic programming and n-person games.
\newblock {\em Artificial Intelligence}, 77:321--358, 1995.

\bibitem{FerrettietalEJOR2018}
V.~Ferretti, I.~Pluchinotta, and A.~Tsouki\`as.
\newblock Studying the generation of alternatives in public policy making
  processes.
\newblock {\em European Journal of Operational Research}, 273:353 -- 363, 2019.
\newblock https://doi.org/10.1016/j.ejor.2018.07.054.

\bibitem{FurnkranzHullermeier2010}
J.~F\"urnkranz and E.~H\"ullermeier.
\newblock {\em Preference Learning}.
\newblock Springer Verlag, Berlin, 2010.

\bibitem{gallo2004}
G.~Gallo.
\newblock Operations research and ethics: Responsibility, sharing and
  cooperation.
\newblock {\em European Journal of Operational Research}, 153(2):468--476,
  2004.

\bibitem{gass2009}
S.~Gass.
\newblock Ethical guidelines and codes in operations research.
\newblock {\em Omega}, 37:1044--1050, 2009.

\bibitem{Gass1994}
{S.I.} Gass.
\newblock On ethics in operational research.
\newblock {\em Journal of the Operational Research Society}, 45:965--966, 1994.

\bibitem{Gigerenzer02}
G.~Gigerenzer.
\newblock {\em Calculated risks: How to know when numbers deceive you}.
\newblock Simon and Schuster, New York, 2002.

\bibitem{GigerenzerTodd99}
G.~Gigerenzer and {P.M.} Todd.
\newblock {\em Simple heuristics that make us smart}.
\newblock Oxford University Press, New York, 1999.

\bibitem{Keeney92}
{R.L}. Keeney.
\newblock {\em Value-{F}ocused {T}hinking. {A} {P}ath to {C}reative {D}ecision
  {M}aking}.
\newblock Harvard {U}niversity {P}ress, Cambridge, 1992.

\bibitem{Kirby06}
{M.W.} Kirby.
\newblock A festering sore: the issue of professionalism in the history of the
  operational research society.
\newblock {\em Journal of the Operational Research Society}, 57:161 -- 172,
  2006.

\bibitem{LandryBanvilleOral96}
M.~Landry, C.~Banville, and M.~Oral.
\newblock Model legitimisation in operational research.
\newblock {\em European Journal of Operational Research}, 92:443--457, 1996.

\bibitem{LandryMalouinOral83}
M.~Landry, {J.L}. Malouin, and M.~Oral.
\newblock Model validation in operations research.
\newblock {\em European Journal of Operational Research}, 14:207--220, 1983.

\bibitem{LeMasson2013}
P.~{Le Masson}, K.~Dorst, and E.~Subrahmanian.
\newblock {Design theory: History, state of the art and advancements}.
\newblock {\em Research in Engineering Design}, 24:97--103, 2013.

\bibitem{LeMenestrelVanWassenhove2009}
M.~{Le Menestrel} and {L.N}. {Van Wassenhove}.
\newblock Ethics in operations research and management sciences: a never-ending
  effort to combine rigor and passion.
\newblock {\em International Journal in Operational Research}, 37:1030--1043,
  2009.

\bibitem{MeinardTsoukiasEJOR2018}
Y.~Meinard and A.~Tsouki\`as.
\newblock On the rationality of decision aiding processes.
\newblock {\em European Journal of Operational Research}, 273:1074 -- 1084,
  2019.

\bibitem{MeinardTsoukias2022}
Y.~Meinard and A.~Tsouki\`as.
\newblock What is legitimate decision support?
\newblock Cahier du {LAMSADE}, 401, PSL, Universit\'e Paris Dauphine, 2022.
\newblock https://arxiv.org/abs/2201.12071.

\bibitem{NarensLuce1990}
L.~Narens and {R.D.} Luce.
\newblock Meaningfulness and invariance.
\newblock In Eatwell J., Milgate M., and Newman P, editors, {\em Time Series
  and Statistics}, pages 140 -- 148. Palgrave Macmillan, London, 1990.

\bibitem{Nott2006}
J.~Nott.
\newblock {\em Extreme Events: A Physical Reconstruction and Risk Assessment}.
\newblock Cambridge University Press, Cambirdge, 2006.

\bibitem{OECD}
OECD.
\newblock Best practices for ensuring scientific integrity and preventing
  misconduct, 2007.
\newblock https://www.oecd.org/science/inno/40188303.pdf.

\bibitem{OrmerodUlrich2013}
{R.J.} Ormerod and W.~Ulrich.
\newblock Operational research and ethics: A literature review.
\newblock {\em European Journal of Operational Research}, 228(2):291--307,
  2013.

\bibitem{ORSA71}
ORSA.
\newblock Guidelines for the practice of operations research.
\newblock {\em Operations Research}, 19:1123--1148, 1971.

\bibitem{Ouerdaneetal2010}
W.~Ouerdane, N.~Maudet, and A.~Tsouki\`as.
\newblock Argumentation theory and decision aiding.
\newblock In J.~Figueira, S.~Greco, and M.~Ehrgott, editors, {\em Trends in
  Multiple Criteria Decision Analysis}, pages 177 -- 208. Springer Verlag, New
  York, 2010.

\bibitem{Rauschmayeretal2009}
F.~Rauschmayer, I.~Kavathatzopoulos, {P.L.} Kunsch, and M.~{Le Menestrel}.
\newblock Why good practice is not enough - ethical challenges for the or
  practitioner.
\newblock {\em Omega}, 37:1089--1099, 2009.

\bibitem{roberts79}
{F.S}. Roberts.
\newblock {\em Measurement theory, with applications to Decision Making,
  Utility and the Social Sciences}.
\newblock Addison-Wesley, Boston, 1979.

\bibitem{Roberts1980}
{F.S.} Roberts.
\newblock On {Luce}'s theory of meaningfulness.
\newblock {\em Philosophy of Science}, 47(3):424--433, 1980.

\bibitem{rosen1book89}
J.~Rosenhead.
\newblock {\em Rational analysis of a problematic world}.
\newblock J. Wiley, New York, 1989.
\newblock 2nd revised edition in 2001.

\bibitem{Rosenhead2006}
J.~Rosenhead.
\newblock {Past, present and future of problem structuring methods}.
\newblock {\em Journal of the Operational Research Society}, 57:759--765, 2006.

\bibitem{RosenheadMitchell1986}
J.~Rosenhead and {G.H.} Mitchell.
\newblock Report of the commission on the future practice of operational
  research.
\newblock {\em Journal of the Operational Research Society}, 37:831--886, 1986.

\bibitem{Smithetal1992}
{B.C.} Smith, {J.F.} Leimkuhler, and {R.M.} Darrow.
\newblock Yield management at {American} {Airlines}.
\newblock {\em Journal on Applied Analytics}, 22:8 -- 31, 1992.

\bibitem{TomasiniWassenhove2009}
R.~Tomasini and L.~{van Wassenhove}.
\newblock {\em Humanitarian Logistics}.
\newblock Palgrave, London, 2009.

\bibitem{TrevorBench-Capon2003}
J.~Trevor and M.~{Bench-Capon}.
\newblock Persuasion in practical argument using value-based argumentation
  frameworks.
\newblock {\em Journal of Logic and Computation}, 13:429 -- 448, 2003.

\bibitem{Tsoukias07aor}
A.~Tsouki\`as.
\newblock On the concept of decision aiding process.
\newblock {\em Annals of Operations Research}, 154:3 -- 27, 2007.

\bibitem{Tsoukias08ejor}
A.~Tsouki\`as.
\newblock From decision theory to decision aiding methodology.
\newblock {\em European Journal of Operational Research}, 187:138 -- 161, 2008.

\bibitem{Tsoukias2021}
A.~Tsouki\`as.
\newblock Social responsibility of algorithms: an overview.
\newblock In J.~Papathanasiou, P.~Zarat\'e, and J.~{Freire de Sousa}, editors,
  {\em {EURO} Working Group on {DSS}}, pages 153 -- 166. Springer Nature, 2021.

\bibitem{vinck1theo92}
{Ph}. Vincke.
\newblock Exploitation of a crisp relation in a ranking problem.
\newblock {\em Theory and Decision}, 32(3):221--240, 1992.

\bibitem{singapore}
WCRIF.
\newblock Singapore statement on research integrity, 2010.
\newblock https://wcrif.org/guidance/singapore-statement.

\bibitem{Wenstop2010}
F.~Wenst{\o}p.
\newblock Operations research and ethics: development trends 1996-2009.
\newblock {\em International Transactions in Operational Research},
  17:413--426, 2010.

\end{thebibliography}

\end{document}